\documentclass[journal]{IEEEtran}
\IEEEoverridecommandlockouts
\usepackage[noadjust]{cite}
\usepackage{amsmath,amssymb,amsfonts}
\usepackage{algorithmic}
\usepackage{graphicx}
\usepackage{textcomp,dsfont}
\usepackage{xcolor,comment}
\usepackage{nicefrac,xfrac}
\usepackage{subcaption}
\usepackage{flushend}

\newtheorem{theorem}{Theorem}
\newtheorem{corollary}{Corollary}
\newtheorem{lemma}{Lemma}

\newtheorem{definition}{Definition}

\def\BibTeX{{\rm B\kern-.05em{\sc i\kern-.025em b}\kern-.08em
    T\kern-.1667em\lower.7ex\hbox{E}\kern-.125emX}}

\begin{document}

\title{A Design Framework that Unifies 6G \\ Modulation Schemes for Double Selectivity}

\author{Nishant Mehrotra$^*$, Sandesh Rao Mattu$^*$, Robert Calderbank~\IEEEmembership{Life Fellow,~IEEE}
\thanks{This work is supported by the National Science Foundation under grants 2342690 and 2148212, in part by funds from federal agency and industry partners as specified in the Resilient \& Intelligent NextG Systems (RINGS) program, and in part by the Air Force Office of Scientific Research under grants FA 8750-20-2-0504 and FA 9550-23-1-0249. \\ The authors are with the Department of Electrical and Computer Engineering, Duke University, Durham, NC, 27708, USA (email: \{nishant.mehrotra,\allowbreak sandesh.mattu,\allowbreak robert.calderbank\}\allowbreak@duke.edu). \\ $*$ denotes equal contribution.}
}


\maketitle
\begin{abstract}
There is significant recent interest in designing new modulation schemes for doubly-selective channels with large delay and Doppler spreads, where legacy modulation schemes based on time-frequency signal representations underperform. Multiple modulation schemes, e.g., in the delay-Doppler, chirp, time-sequency, and other domains, have been proposed in the literature for this purpose, with varying implementation details. In this letter, we establish that all previously proposed modulation schemes for doubly-selective signaling are instances of a \emph{single family} of complex Hadamard-modulated pulse trains. When the delay and Doppler spread of the doubly-selective channel is limited to a certain support, all modulation schemes in this waveform family offer \emph{equivalent}, full diversity achieving performance with no symbol fading and low channel estimation overhead. The existence of this waveform family also enables \emph{flexible, multi-waveform co-existence} -- allowing a common transceiver architecture to generate multiple waveforms in the family, that may each be flexibly allocated to different users and services. 
\end{abstract}


\section{Introduction}
\label{sec:intro}

\IEEEPARstart{6}{G} wireless environments are expected to feature significant delay and Doppler selectivity with the move towards non-terrestrial networks and higher carrier frequencies. In such doubly-selective environments, legacy modulation schemes based on time-frequency signal representations (such as OFDM -- orthogonal frequency division multiplexing~\cite{Ebert1971_ofdm,Bingham1990_ofdm}) do not perform well, and there is growing interest in new modulation schemes for these environments~\cite{Bemani2023_afdm,Cho2025_dftpfdma,Zhao2016_ocdm,Yuan2022_oddm,Viterbo2021_otsm,bitspaper1,bitspaper2,otfs_book,Hadani2017_mcotfs,Chockalingam2019_mcotfs_div,Hanzo2024_otsm_amp,Hanzo2025_afdm_gen,Yuan2024_oddm_gen,Yuan2026_divrot}. 

Examples of such modulation schemes include Zak-OTFS (orthogonal time frequency space modulation)~\cite{bitspaper1,bitspaper2,otfs_book} and ODDM (orthogonal delay-Doppler division multiplexing)~\cite{Yuan2022_oddm,Yuan2024_oddm_gen} that modulate information in the \emph{delay-Doppler} (DD) domain, AFDM (affine frequency division multiplexing)~\cite{Bemani2023_afdm,Cho2025_dftpfdma,Zhao2016_ocdm,Hanzo2025_afdm_gen} that modulates information in the \emph{chirp} domain, and OTSM (orthogonal time sequency division multiplexing)~\cite{Viterbo2021_otsm,Hanzo2024_otsm_amp} that modulates information in the \emph{time-sequency} domain.

In this letter, we unify the literature on this topic by establishing that these examples all correspond to different instances of a \emph{single family} $\Psi$ of waveforms, namely the family of \emph{complex Hadamard-modulated pulse trains}. The family comprises unitary transformations of periodic pulse trains modulated by complex Hadamard matrices. Different choices of unitary transformations and complex Hadamard matrices generate the various modulations proposed in the literature, including Zak-OTFS, ODDM, AFDM\footnote{AFDM has been shown to be equivalent to OCDM (orthogonal chirp division multiplexing)~\cite{Zhao2016_ocdm} and DFT-p-FDMA (discrete Fourier transform phase rotated \& permuted frequency division multiple access)~\cite{Cho2025_dftpfdma} for specific AFDM parameter choices, hence we do not consider those modulations separately.} and OTSM. Table~\ref{tab:prior_work} summarizes various modulations studied in this letter.

When the doubly-selective channel has delay and Doppler spread confined to a certain support, \emph{all} modulation schemes belonging to the defined waveform family $\Psi$ do not undergo symbol fading (termed \emph{non-selectivity}) and enable channel estimation with only a single pilot transmission (termed \emph{predictability}). Moreover, all modulation schemes in the family\footnote{\textcolor{black}{The conventional multicarrier implementation of OTFS (MC-OTFS) as well as time-frequency modulations like OFDM are not part of the defined waveform family, and have been shown to be selective and non-predictable in~\cite{otfs_book,bitspaper1}. Moreover, achieving full time-frequency / delay-Doppler diversity with OFDM and MC-OTFS requires constellation rotation~\cite{Chockalingam2019_mcotfs_div,Yuan2026_divrot}.}} provide \emph{equivalent}, full diversity achieving performance.

In addition to generalizing previously known waveforms, our framework also allows \emph{multiplexing} multiple waveforms in a single transmission. This capability enables flexibly allocating different waveforms to users with heterogeneous service requirements, and hence may be of interest in next-generation ($6$G) networks for multi-user, multi-service network operation.

\begin{table*}
    \centering
    \caption{Different modulation schemes for doubly-selective signaling using a frame of bandwidth $B$ and interval $T$, over a channel with delay spread $\tau_{\max}$. $\Psi$ denotes the family of complex Hadamard-modulated pulse trains defined in Section~\ref{sec:family}. $O(a) / O(b(k))$ indicates complexity $O(a)$ (resp. $O(b(k))$) using a non-iterative (resp. iterative) equalizer across $k$ iterations.} 
    {
    \setlength{\tabcolsep}{2.25pt}
    \renewcommand{\arraystretch}{1.25}
    \begin{tabular}{|c|c|c|c|c|c|}
         \hline
         Modulation Scheme & \textcolor{black}{Member of $\Psi$?} & \textcolor{black}{TX \& RX Complexity} & \textcolor{black}{Equalization Complexity} & \textcolor{black}{Diversity Order} & \textcolor{black}{Per-Symbol Cyclic Prefix Overhead} \\ 
         \hline
         OFDM~\cite{Ebert1971_ofdm,Bingham1990_ofdm} & \textcolor{black}{$\times$} & \textcolor{black}{$O(BT\log B)$} & \textcolor{black}{$O(BT) / O(kBT)$} & \textcolor{black}{$1$} & \textcolor{black}{$\lceil B\tau_{\max} \rceil$} \\
         \textcolor{black}{MC-OTFS~\cite{Hadani2017_mcotfs,otfs_book,Chockalingam2019_mcotfs_div}} & \textcolor{black}{$\times$} & \textcolor{black}{$O(BT\log BT)$} & \textcolor{black}{$O(BT\log BT) / O(kBT)$} & \textcolor{black}{$1$} & \textcolor{black}{$\lceil B\tau_{\max} \rceil$} \\
         AFDM\textcolor{black}{~\cite{Bemani2023_afdm,Cho2025_dftpfdma,Zhao2016_ocdm,Hanzo2025_afdm_gen}} & \textcolor{black}{\checkmark} & \textcolor{black}{$O(BT\log B)$} & \textcolor{black}{$O\big((BT)^{3}\big) / O(kBT)$} & \textcolor{black}{Full} & \textcolor{black}{$\lceil B\tau_{\max} \rceil$} \\ 
         ODDM\textcolor{black}{~\cite{Yuan2022_oddm,Yuan2024_oddm_gen}} & \textcolor{black}{\checkmark} & \textcolor{black}{$O(BT\log T)$} & \textcolor{black}{$O\big((BT)^{3}\big) / O(kBT)$} & \textcolor{black}{Full} & \textcolor{black}{$0$} \\ 
         OTSM\textcolor{black}{~\cite{Viterbo2021_otsm,Hanzo2024_otsm_amp}} & \textcolor{black}{\checkmark} & \textcolor{black}{$O(BT\log T)$} & \textcolor{black}{$O\big((BT)^{3}\big) / O(kBT)$} & \textcolor{black}{Full} & \textcolor{black}{$0$} \\
         Zak-OTFS~\cite{bitspaper1,bitspaper2,otfs_book} & \textcolor{black}{\checkmark} & \textcolor{black}{$O(BT\log T)$} & \textcolor{black}{$O\big((BT)^{3}\big) / O(kBT)$} & \textcolor{black}{Full} & \textcolor{black}{$0$} \\
         \hline
    \end{tabular}
    }
    \label{tab:prior_work}
\end{table*}

\textit{Notation:} $x$ denotes a complex scalar, $\mathbf{x}$ denotes a vector with $n$th entry $\mathbf{x}[n]$, and $\mathbf{X}$ denotes a matrix with $(n,m)$th entry $\mathbf{X}[n,m]$. $(\cdot)^{\ast}$ denotes complex conjugate, $(\cdot)^{\top}$ denotes transpose, $(\cdot)^{\mathsf{H}}$ denotes complex conjugate transpose. $\mathbb{Z}$ denotes the set of integers, $\mathbb{Z}_{N}$ the set of integers modulo $N$, and $(\cdot)_{{}_{N}}$ denotes the value modulo $N$. $\lfloor \cdot \rfloor$ and $\lceil \cdot \rceil$ denote the floor and ceiling functions. $a \odot b$ and $(a,b)$ respectively denote the bitwise dot product and greatest common divisor of two integers $a,b$. $\delta(\cdot)$ denotes the delta function, $\delta[\cdot]$ denotes the Kronecker delta function, $\mathds{1}{\{\cdot\}}$ denotes the indicator function, $\mathbf{I}_{N}$ denotes the $N \times N$ identity matrix, and $\mathbf{e}_{n}$ is the standard basis vector with value $1$ at location $n$ and zero elsewhere. 

\section{Signaling over Doubly-Selective Channels}
\label{sec:prelim}

In this Section, we derive a general system model for communication over doubly-selective channels and apply it to model different modulation schemes proposed in the literature. 

The continuous linear time varying system model for communication over a doubly-selective channel is~\cite{Bello1963_ltv,bitspaper1,bitspaper2,otfs_book}:
\begin{align}
    \label{eq:prelim1}
    y(t) &= \iint h(\tau,\nu) x(t-\tau) e^{j2\pi\nu(t-\tau)} d\tau d\nu + w(t),
\end{align}
where $x(t)$ (resp. $y(t)$) denotes the transmit (resp. receive) waveform in continuous time, $w(t)$ denotes the additive noise at the receiver, and $h(\tau,\nu)$ represents the channel spreading function in delay $\tau$ and Doppler $\nu$. 

We assume communication occurs over a finite bandwidth $B$ and time interval $T$, where we assume the time-bandwidth product $BT$ is an integer. Hence, we consider the discrete time version of the system model in~\eqref{eq:prelim1}, with \textcolor{black}{$BT$ samples} of the transmit and receive waveforms at integer multiples of the delay resolution $\nicefrac{1}{B}$ and limited to duration $T$~\cite{Mehrotra2025_EURASIP,Mattu2025_npj}:
\begin{align}
    \label{eq:prelim2}
    \mathbf{y}[n] &= \sum_{k,l \in \mathbb{Z}_{BT}} \mathbf{h}[k,l] \mathbf{x}[(n-k)_{{}_{BT}}] e^{\frac{j2\pi}{BT}l(n-k)} + \mathbf{w}[n],
\end{align}
where $n \in \mathbb{Z}_{BT}$ denotes the sampling index ($n = \lfloor Bt \rfloor$ for $0 \leq t \leq T$), $\mathbf{w}$ denotes noise samples, and $\mathbf{h}[k,l] = h\big(\nicefrac{k}{B},\nicefrac{l}{T}\big)$ is the channel spreading function sampled at \textcolor{black}{$BT \times BT$} integer multiples of the delay / Doppler resolutions. 

Note that the transmit and receive waveforms $\mathbf{x}$ and $\mathbf{y}$ in~\eqref{eq:prelim2} are $BT$-periodic sequences. Hence, $BT$ information symbols can be transmitted via a $BT$-dimensional orthonormal basis:
\begin{align}
    \label{eq:prelim3}
    \mathbf{x}[n] &= \sum_{i \in \mathbb{Z}_{BT}} \mathbf{s}[i] \boldsymbol{\phi}_{i}[n],
\end{align}
where $\mathbf{s}$ denotes the $BT$-length vector of information symbols and $\boldsymbol{\phi}$ is an orthonormal basis with $BT$ elements, each of length $BT$. Substituting~\eqref{eq:prelim3} in~\eqref{eq:prelim2} and projecting the receive waveform $\mathbf{y}$ on the basis $\boldsymbol{\phi}$, we obtain the system model~\cite{Mattu2025_npj}:
\begin{align}
    \label{eq:prelim4}
    \mathbf{r}[f] = &\sum_{n \in \mathbb{Z}_{BT}} \boldsymbol{\phi}_{f}^{*}[n] \mathbf{y}[n] \nonumber \\
    = &\sum_{i \in \mathbb{Z}_{BT}} \mathbf{s}[i] \bigg(\sum_{k,l \in \mathbb{Z}_{BT}} e^{-\frac{j2\pi}{BT}lk} \mathbf{h}[k,l] \nonumber \\ \times &\sum_{n \in \mathbb{Z}_{BT}} \boldsymbol{\phi}_{f}^{*}[n] \boldsymbol{\phi}_{i}[(n-k)_{{}_{BT}}] e^{\frac{j2\pi}{BT}ln}\bigg) + \mathbf{v}[f] \nonumber \\
    = &\sum_{i \in \mathbb{Z}_{BT}} \mathbf{s}[i] \mathbf{G}[f,i] + \mathbf{v}[f],
\end{align}
where $\mathbf{v}[f] = \sum_{n \in \mathbb{Z}_{BT}} \boldsymbol{\phi}_{f}^{*}[n] \mathbf{w}[n]$ is the projection of noise samples on the \textcolor{black}{$f$th basis element in $\boldsymbol{\phi}$}. On vectorizing~\eqref{eq:prelim4}:
\begin{align}
    \label{eq:prelim5}
    \mathbf{r} &= \mathbf{G} \mathbf{s} + \mathbf{v},
\end{align}
where $\mathbf{G}$ denotes the equivalent $BT \times BT$ channel matrix.

Recovering the transmitted information symbols $\mathbf{s}$ requires knowledge of the channel matrix $\mathbf{G}$, or equivalently, the sampled channel spreading function $\mathbf{h}[k,l]$. The sampled channel spreading function can be estimated by transmitting a known pilot symbol and computing the \emph{cross-ambiguity function}~\cite{Calderbank2025_isac} between the received and transmitted waveforms: 
\begin{align}
    \label{eq:prelim6}
    \widehat{\mathbf{h}}[k,l] &= \mathbf{A}_{\mathbf{y},\mathbf{x}}[k,l] \nonumber \\
    &= \sum_{n \in \mathbb{Z}_{BT}} \mathbf{y}[n] \mathbf{x}^{*}[(n-k)_{{}_{BT}}]e^{-\frac{j2\pi}{BT}l(n-k)},
\end{align}
which has been shown to be the maximum likelihood estimate in~\cite{Calderbank2025_isac}. Subsequently, the matrix $\mathbf{G}$ is estimated using~\eqref{eq:prelim4} and used to recover the information symbols $\mathbf{s}$, e.g., via the minimum mean squared error (MMSE) estimator~\cite{otfs_book}.

\subsection{Examples of Modulation Schemes for Double Selectivity}
\label{subsec:prelim_mod}

We now present examples of various modulation schemes proposed in the literature using the system model in~\eqref{eq:prelim5}. To that end, we assume a time-frequency (equivalently, delay-Doppler) frame with $M$ subcarriers (delay bins) spaced apart at $\Delta f$, such that $B = M \Delta f$, and $N$ symbols (Doppler bins) of duration $\nicefrac{1}{\Delta f}$ each, such that $T = \nicefrac{N}{\Delta f}$. Hence, $BT = MN$.

\subsubsection{OFDM}
\label{subsubsec:prelim_ofdm}

The basis element in OFDM is~\cite{Ebert1971_ofdm,Bingham1990_ofdm}:
\begin{align}
    \label{eq:ofdm1}
    \boldsymbol{\phi}_{i}[n] = \frac{1}{\sqrt{M}} e^{\frac{j2\pi}{M}in} \mathds{1}\big\{\lfloor\nicefrac{n}{M}\rfloor = \lfloor\nicefrac{i}{M}\rfloor\big\}.
\end{align}

\subsubsection{AFDM}
\label{subsubsec:prelim_afdm}

The basis element in AFDM is~\cite{Bemani2023_afdm,Cho2025_dftpfdma,Hanzo2025_afdm_gen}:
\begin{align}
    \label{eq:afdm1}
    \boldsymbol{\phi}_{i}[n] = \frac{1}{\sqrt{MN}}e^{j2\pi\big(c_1n^2+c_2i^2+\frac{ni}{MN}\big)},
\end{align}
where $c_1, c_2 \in \mathbb{Z}$. The AFDM basis specializes to OCDM~\cite{Bemani2023_afdm} when $c_1 = c_2 = \nicefrac{1}{2MN}$ and to DFT-p-FDMA~\cite{Cho2025_dftpfdma} when $c_1 = c_2 = \nicefrac{\Delta}{MN}$, where $(\Delta,MN) = 1$.

\subsubsection{ODDM}
\label{subsubsec:prelim_oddm}

The basis element in ODDM is~\cite{Yuan2022_oddm,Yuan2024_oddm_gen}:
\begin{align}
    \label{eq:oddm1}
    \boldsymbol{\phi}_{i}[n] = \frac{1}{\sqrt{N}} e^{\frac{j2\pi}{N}{\lfloor\nicefrac{i}{M}\rfloor} \lfloor\nicefrac{n}{M}\rfloor} \mathds{1}\big\{n \equiv i \bmod{M}\big\}.
\end{align}

\subsubsection{OTSM}
\label{subsubsec:prelim_otsm}

The basis element in OTSM is~\cite{Viterbo2021_otsm,Hanzo2024_otsm_amp}:
\begin{align}
    \label{eq:otsm1}
    \boldsymbol{\phi}_{i}[n] = \frac{1}{\sqrt{N}} (-1)^{\lfloor\nicefrac{i}{M}\rfloor \odot \lfloor\nicefrac{n}{M}\rfloor} \mathds{1}\big\{n \equiv i \bmod{M}\big\},
\end{align}
where $\odot$ denotes the bitwise dot product.

\subsubsection{Zak-OTFS}
\label{subsubsec:prelim_zak}

The basis element in Zak-OTFS is~\cite{bitspaper1,bitspaper2,otfs_book}:
\begin{align}
    \label{eq:zakotfs1}
    \boldsymbol{\phi}_{i}[n] &= \frac{1}{\sqrt{N}} \sum_{d \in \mathbb{Z}} e^{j\frac{2\pi}{N} d \lfloor \nicefrac{i}{M} \rfloor} \delta[n-(i)_{{}_{M}}-dM] \nonumber \\
    &\textcolor{black}{= \frac{1}{\sqrt{N}} e^{\frac{j2\pi}{N}{\lfloor\nicefrac{i}{M}\rfloor} \lfloor\nicefrac{n}{M}\rfloor} \mathds{1}\big\{n \equiv i \bmod{M}\big\},}
\end{align}
termed \emph{pulsone} due to its structure of a pulse train modulated by a tone, \textcolor{black}{and coincides with the ODDM basis element in~\eqref{eq:oddm1}.}

\section{A Waveform Family for Double Selectivity}
\label{sec:family}

In this Section, we establish that the modulations described in Section~\ref{subsec:prelim_mod} all belong to a \emph{single family} of complex Hadamard-modulated pulse train waveforms. Under benign channel conditions, \emph{all} waveforms in the defined family are non-selective and predictable, thus offer equivalent, full diversity achieving performance. We also highlight potential applications of our framework to multi-waveform co-existence.

\subsection{Family of Complex Hadamard-Modulated Pulse Trains}
\label{subsec:family_complex_hadamard}

Consider the following family of $MN$-periodic waveforms:
\begin{align}
    \label{eq:sys1}
    \Psi = \bigg\{\mathbf{U} \boldsymbol{\phi}_{i}:~&\boldsymbol{\phi}_{i}[n] = \frac{1}{\sqrt{N}} \mathbf{H}_{(i)_{{}_{M}}}[\lfloor \nicefrac{n}{M} \rfloor, \lfloor \nicefrac{i}{M} \rfloor] \nonumber \\ &\times \mathds{1}\big\{n \equiv i \bmod{M}\big\}, i,n \in \mathbb{Z}_{MN} \bigg\},
\end{align}
where $\mathbf{U}$ is an arbitrary $MN \times MN$ unitary matrix, and $\mathbf{H}_{x}$ denotes an $N \times N$ \emph{complex Hadamard matrix} parameterized by index $x \in \mathbb{Z}_{M}$ and with $(p,q)$th entry $\mathbf{H}_{x}[p,q]$.

\begin{definition}[\cite{Butson1962_hadamard,Tadej2006_hadamard,Banica2021_hadamard1,Banica2024_hadamard2}]
    \label{def:complex_Hadamard}
    An $N \times N$ complex Hadamard matrix $\mathbf{H}$ is a square matrix with unimodular entries: $|\mathbf{H}[p,q]| = 1$ for all $p,q \in \mathbb{Z}_{N}$, and pairwise orthogonal rows \& columns: $\mathbf{H}^{\mathsf{H}} \mathbf{H} = \mathbf{H} \mathbf{H}^{\mathsf{H}} = N \mathbf{I}_{N}$.
\end{definition}

Examples of complex Hadamard matrices include the discrete Fourier transform (DFT) matrix and its inverse (IDFT), the Walsh matrix, etc. Different choices of the unitary matrix $\mathbf{U}$ and the complex Hadamard matrix $\mathbf{H}_{(i)_{{}_{M}}}$ in~\eqref{eq:sys1} result in different modulations (except OFDM) from Section~\ref{subsec:prelim_mod}.

\subsubsection{Zak-OTFS \& ODDM}
\label{subsubsec:family_zak}

Choose $\mathbf{U} = \mathbf{I}_{MN}$ and $\mathbf{H}_{(i)_{{}_{M}}}$ as the IDFT matrix with entry $\mathbf{H}_{(i)_{{}_{M}}}[p,q] = e^{j\frac{2\pi}{N}pq}$.

\subsubsection{OTSM}
\label{subsubsec:family_otsm}

Choose $\mathbf{U} = \mathbf{I}_{MN}$ and $\mathbf{H}_{(i)_{{}_{M}}}$ as the Walsh matrix with entry $\mathbf{H}_{(i)_{{}_{M}}}[p,q] = (-1)^{p \odot q}$.

\subsubsection{AFDM}
\label{subsubsec:family_afdm}

Choose $\mathbf{U}$ as the generalized discrete affine Fourier transform (GDAFT) matrix from~\cite{Mehrotra2025_WCLSpread,Mehrotra2025_EURASIP} and $\mathbf{H}_{(i)_{{}_{M}}}$ as the IDFT matrix with entry $\mathbf{H}_{(i)_{{}_{M}}}[p,q] = e^{j\frac{2\pi}{N}pq}$.

\subsection{Properties of Modulation Schemes in $\Psi$}
\label{subsec:family_prop}

We now derive key properties satisfied by every modulation scheme belonging to the waveform family $\Psi$. Specifically, we derive conditions under which the channel estimate $\widehat{\mathbf{h}}[k,l]$ obtained via~\eqref{eq:prelim6} matches the true channel spreading function $\mathbf{h}[k,l]$ regardless of which element $\boldsymbol{\phi}_{i},~i\in\mathbb{Z}_{MN}$ in the modulation basis is transmitted as a pilot symbol.

\begin{lemma}[\cite{Calderbank2025_isac,Mehrotra2025_EURASIP}]
    \label{lmm:pred}
    Let $\mathcal{S} = \big\{(k,l):|\mathbf{h}[k,l]|\neq0\big\}$ denote the channel support corresponding to spreading function $\mathbf{h}[k,l]$. Define $\mathcal{K}_{\mathcal{S}} = \big\{(k',l'):\mathcal{S}\cap\big(\mathcal{S}+(k',l')\big)\neq\emptyset\big\}$ as the set of DD locations where translates of the channel support overlap with $\mathcal{S}$. Then, channel estimates via~\eqref{eq:prelim6} always equal $\mathbf{h}[k,l]$ for all $(k,l)\in\mathcal{S},i\in\mathbb{Z}_{MN}$ in a modulation basis with:
    \begin{align*}
        \mathbf{A}_{\boldsymbol{\phi}_{i}}[0,0] = 1,\mathbf{A}_{\boldsymbol{\phi}_{i}}[k',l'] = 0~\text{for}~(k',l')\in\mathcal{K}_{\mathcal{S}}\setminus\{(0,0)\},
    \end{align*}
    where $\mathbf{A}_{\mathbf{x}}[k,l]$ denotes the self-ambiguity of $\mathbf{x}$ ($\mathbf{y} = \mathbf{x}$ in~\eqref{eq:prelim6}).
\end{lemma}

\begin{IEEEproof}
    See~\cite[Section V-B]{Calderbank2025_isac} for the proof.
\end{IEEEproof}

\begin{figure}
    \centering
    \includegraphics[width=0.92\linewidth]{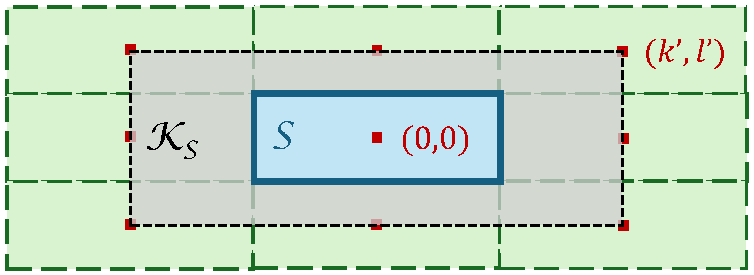}
    \caption{\textcolor{black}{For predictability (Lemma~\ref{lmm:pred}) with a rectangular channel support $\mathcal{S}$, the self-ambiguity $\mathbf{A}_{\boldsymbol{\phi}_{i}}[k',l']$ should have only $1$ non-zero value at $(k',l')=(0,0)$ within the region $\mathcal{K}_{\mathcal{S}}$ (gray).}}
    \label{fig:cryst}
\end{figure}

We term the above condition \emph{predictability} since it implies that the channel response to any basis element $i\in\mathbb{Z}_{MN}$ can be exactly predicted from the (noiseless) measurements obtained for a given basis element. Fig.~\ref{fig:cryst} geometrically illustrates the condition for a simple example of a rectangular support $\mathcal{S}$.

\begin{corollary}
    \label{corr:nonsel}
    When predictability as per Lemma~\ref{lmm:pred} holds, then the constituent carriers in the modulation basis do not undergo symbol fading\footnote{\textcolor{black}{Assuming $\mathbf{s} = c \cdot \mathbf{e}_{i}$ for unit-energy symbol $c$ on standard basis vector / carrier $i$ and calculating $\mathbf{r}^{\mathsf{H}} \mathbf{r} = \mathbf{s}^{\mathsf{H}} \mathbf{G}^{\mathsf{H}}\mathbf{G} \mathbf{s} = |c|^{2} \mathbf{e}_{i}^{\mathsf{H}} \mathbf{G}^{\mathsf{H}}\mathbf{G} \mathbf{e}_{i} = (\mathbf{G}^{\mathsf{H}}\mathbf{G})[i,i]$.}}, i.e., for all $i,j\in\mathbb{Z}_{MN}$:
    \begin{align*}
        (\mathbf{G}^{\mathsf{H}}\mathbf{G})[i,i] = (\mathbf{G}^{\mathsf{H}}\mathbf{G})[j,j] = \sum_{k,l\in\mathcal{S}}|\mathbf{h}[k,l]|^{2}.
    \end{align*}
\end{corollary}


\begin{IEEEproof}
    Evaluating $\mathbf{G}^{\mathsf{H}}\mathbf{G}$ via~\eqref{eq:prelim4} (with $BT = MN$):
    \begin{align}
        \label{eq:non_fading2}
        (\mathbf{G}^{\mathsf{H}}\mathbf{G})[i,i] = &\sum_{k_1,k_2} \sum_{l_1,l_2} \mathbf{h}^{*}[k_1,l_1] \mathbf{h}[k_2,l_2] e^{\frac{j2\pi}{MN}(k_1 l_1 - k_2 l_2)} \nonumber \\ \times &\sum_{n_1,n_2} \bigg(\sum_{f=0}^{MN-1} \boldsymbol{\phi}_{f}[n_1] \boldsymbol{\phi}_{f}^{*}[n_2]\bigg) e^{\frac{j2\pi}{MN}(l_2 n_2 - l_1 n_1)} \nonumber \\ \times &\boldsymbol{\phi}_{i}[(n_2-k_2)_{{}_{MN}}] \boldsymbol{\phi}_{i}^{*}[(n_1-k_1)_{{}_{MN}}].
    \end{align}
    
    For an orthonormal basis, by definition, the summation over $f$ evaluates to $\delta[n_1-n_2]$. Therefore, $n_1 = n_2 = n$ and hence:
    \begin{align}
        \label{eq:non_fading3}
        &(\mathbf{G}^{\mathsf{H}}\mathbf{G})[i,i] = \sum_{k_1,k_2} \sum_{l_1,l_2} \mathbf{h}^{*}[k_1,l_1] \mathbf{h}[k_2,l_2] e^{\frac{j2\pi}{MN}(k_1 l_1 - k_2 l_2)} \nonumber \\ &~~~\times \sum_{n} \boldsymbol{\phi}_{i}[(n-k_2)_{{}_{MN}}] \boldsymbol{\phi}_{i}^{*}[(n-k_1)_{{}_{MN}}] e^{\frac{j2\pi}{MN}(l_2 - l_1) n}.
    \end{align}

    On substituting $\bar{n} = (n-k_2)_{{}_{MN}}$ and simplifying:
    \begin{align}
        \label{eq:non_fading4}
        (\mathbf{G}^{\mathsf{H}}\mathbf{G})[i,i] = \sum_{k_1,k_2} \sum_{l_1,l_2} &\mathbf{h}^{*}[k_1,l_1] \mathbf{h}[k_2,l_2] e^{\frac{j2\pi}{MN}l_2(k_1 - k_2)} \nonumber \\ \times &\mathbf{A}_{\boldsymbol{\phi}_{i}}[k_1-k_2,l_1-l_2].
    \end{align}

    Since $(k_1,l_1),(k_2,l_2)\in\mathcal{S}$, Lemma~\ref{lmm:pred} implies $k_1=k_2$ and $l_1=l_2$; hence, $(\mathbf{G}^{\mathsf{H}}\mathbf{G})[i,i] = \sum_{k,l\in\mathcal{S}}|\mathbf{h}[k,l]|^{2}$.
\end{IEEEproof}

\begin{figure*}
    \centering
    \begin{subfigure}{\linewidth}
        \includegraphics[width=\textwidth]{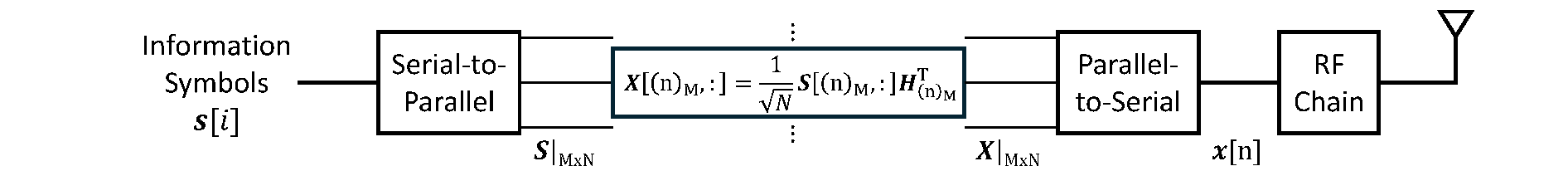}
    \caption{\textcolor{black}{Transmitter block diagram.}}
    \label{fig:txrx_tx}
    \end{subfigure}
    \vspace{0.5em}
    \begin{subfigure}{\linewidth}
        \includegraphics[width=\textwidth]{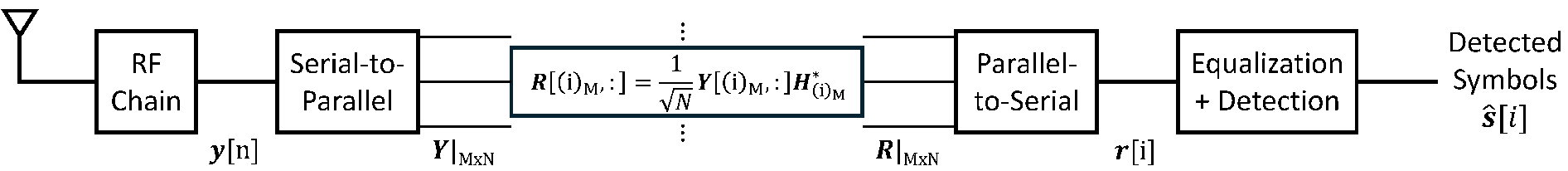}
    \caption{\textcolor{black}{Receiver block diagram.}}
    \label{fig:txrx_rx}
    \end{subfigure}
    \caption{\textcolor{black}{Transmitter and receiver block diagrams implementing matrix-vector form of~\eqref{eq:tx1} and~\eqref{eq:rx1} (see footnote~\ref{ftnt:txrx}).}}
    \label{fig:txrx}
\end{figure*}

We term the above condition \emph{non-selectivity}. We now show that all modulations in $\Psi$ are predictable and non-selective. For simplicity, we focus on the case $\mathbf{U} = \mathbf{I}_{MN}$, and analogous results can be similarly derived for specific choices\footnote{\textcolor{black}{See~\cite{Mehrotra2025_WCLSpread,Mehrotra2025_EURASIP} for an example with $\mathbf{U}$ as the GDAFT matrix.}} of $\mathbf{U}$.

\begin{theorem}
    \label{thm:family_pred}
    For $\mathbf{U} = \mathbf{I}_{MN}$, all modulation schemes in $\Psi$ are predictable and non-selective when the channel support $\mathcal{S}$ satisfies $\mathcal{S}\cap\bigg(\bigcup_{(a,b)\neq(0,0)}\big(\mathcal{S}+(aM,bN)\big)\bigg)=\emptyset$.
\end{theorem}


\begin{IEEEproof}
    Computing the self-ambiguity of $\boldsymbol{\phi}_{i}$ in~\eqref{eq:sys1}:
    \begin{align*}
        \mathbf{A}_{\boldsymbol{\phi}_{i}}[k,l] &= \sum_{n \in \mathbb{Z}_{MN}} \boldsymbol{\phi}_{i}[n] \boldsymbol{\phi}_{i}^{*}[(n-k)_{{}_{MN}}]e^{-\frac{j2\pi}{MN}l(n-k)} \nonumber \\
        &= \frac{1}{N} \sum_{n \in \mathbb{Z}_{MN}} \mathbf{H}_{(i)_{{}_{M}}}[\lfloor \nicefrac{n}{M} \rfloor, \lfloor \nicefrac{i}{M} \rfloor] e^{-\frac{j2\pi}{MN}l(n-k)} \nonumber \\ &\times \mathbf{H}_{(i)_{{}_{M}}}^{*}[\lfloor \nicefrac{(n-k)_{{}_{MN}}}{M} \rfloor, \lfloor \nicefrac{i}{M} \rfloor] \mathds{1}\big\{n \equiv i \bmod{M}\big\} \nonumber \\ &\times \mathds{1}\big\{(n-k)_{{}_{MN}} \equiv i \bmod{M}\big\}.
    \end{align*}

    The indicator function implies $k \equiv 0 \bmod{M}$ and $n \equiv i \bmod{M}$. Therefore, on substituting $k = a M,~a \in \mathbb{Z}_{N}$ and $n = (i)_{{}_{M}} + \Delta M,~\Delta \in \mathbb{Z}_{N}$:
    \begin{align*}
        \mathbf{A}_{\boldsymbol{\phi}_{i}}[k,l] &= \frac{e^{-\frac{j2\pi}{N}l(i)_{{}_{M}}}}{N} \sum_{a \in \mathbb{Z}_{N}} \sum_{\Delta \in \mathbb{Z}_{N}} \mathbf{H}_{(i)_{{}_{M}}}^{*}[(\Delta-a)_{{}_{N}}, \lfloor \nicefrac{i}{M} \rfloor] \nonumber \\ &~~~~~~~~\times \mathbf{H}_{(i)_{{}_{M}}}[\Delta, \lfloor \nicefrac{i}{M} \rfloor] e^{-\frac{j2\pi}{N}l(\Delta-a)} \mathds{1}\big\{k = aM\big\} \nonumber \\
        &= \frac{e^{-\frac{j2\pi}{N}l(i)_{{}_{M}}}}{N} \bigg[\sum_{\Delta \in \mathbb{Z}_{N}} \underbrace{\big|\mathbf{H}_{(i)_{{}_{M}}}[\Delta, \lfloor \nicefrac{i}{M} \rfloor]\big|^{2}}_{1~\text{(c.f. Definition~\ref{def:complex_Hadamard})}} e^{-\frac{j2\pi}{N}l\Delta} \nonumber \\ &\times\mathds{1}\big\{k = 0\big\} + \sum_{a = 1}^{N-1} \sum_{\Delta \in \mathbb{Z}_{N}} \mathbf{H}_{(i)_{{}_{M}}}^{*}[(\Delta-a)_{{}_{N}}, \lfloor \nicefrac{i}{M} \rfloor] \nonumber \\ &~~~~~~~\times \mathbf{H}_{(i)_{{}_{M}}}[\Delta, \lfloor \nicefrac{i}{M} \rfloor] e^{-\frac{j2\pi}{N}l(\Delta-a)} \mathds{1}\big\{k = aM\big\} \bigg] \nonumber \\
        &= \frac{e^{-\frac{j2\pi}{N}l(i)_{{}_{M}}}}{N} \bigg[N \mathds{1}\big\{k = 0\big\} \mathds{1}\big\{l \equiv 0 \bmod{N}\big\} \nonumber \\
        &+ \sum_{a = 1}^{N-1} \sum_{\Delta \in \mathbb{Z}_{N}} \mathbf{H}_{(i)_{{}_{M}}}^{*}[(\Delta-a)_{{}_{N}}, \lfloor \nicefrac{i}{M} \rfloor] e^{-\frac{j2\pi}{N}l(\Delta-a)} \nonumber \\ &~~~~~~~~~~~~~~~~~~~~~\times \mathbf{H}_{(i)_{{}_{M}}}[\Delta, \lfloor \nicefrac{i}{M} \rfloor] \mathds{1}\big\{k = aM\big\} \bigg],
    \end{align*}
    where the final expression is due to the sum of $N$th roots of unity. At Doppler indices $l \equiv 0 \bmod{N}$ corresponding to the first term, the phase terms $e^{-\frac{j2\pi}{N}l(i)_{{}_{M}}}$ vanish; hence:
    \begin{align*}
        \mathbf{A}_{\boldsymbol{\phi}_{i}}[k,l] &= \mathds{1}\big\{k = 0\big\} \mathds{1}\big\{l \equiv 0 \bmod{N}\big\} \nonumber \\
        &~~+ \frac{e^{-\frac{j2\pi}{N}l(i)_{{}_{M}}}}{N} \sum_{a = 1}^{N-1} \sum_{\Delta \in \mathbb{Z}_{N}} \mathbf{H}_{(i)_{{}_{M}}}^{*}[(\Delta-a)_{{}_{N}}, \lfloor \nicefrac{i}{M} \rfloor] \nonumber \\ &~~~~~~~~\times e^{-\frac{j2\pi}{N}l(\Delta-a)} \mathbf{H}_{(i)_{{}_{M}}}[\Delta, \lfloor \nicefrac{i}{M} \rfloor]  \mathds{1}\big\{k = aM\big\}.
    \end{align*}

   Since $\mathcal{S}\cap\bigg(\bigcup_{(a,b)\neq(0,0)}\big(\mathcal{S}+(aM,bN)\big)\bigg)=\emptyset$, the condition in Lemma~\ref{lmm:pred} is satisfied for every $\boldsymbol{\phi}_{i} \in \Psi$.
\end{IEEEproof}

\begin{figure*}
    \centering
    \begin{subfigure}{0.43\linewidth}
    \includegraphics[width=\textwidth]{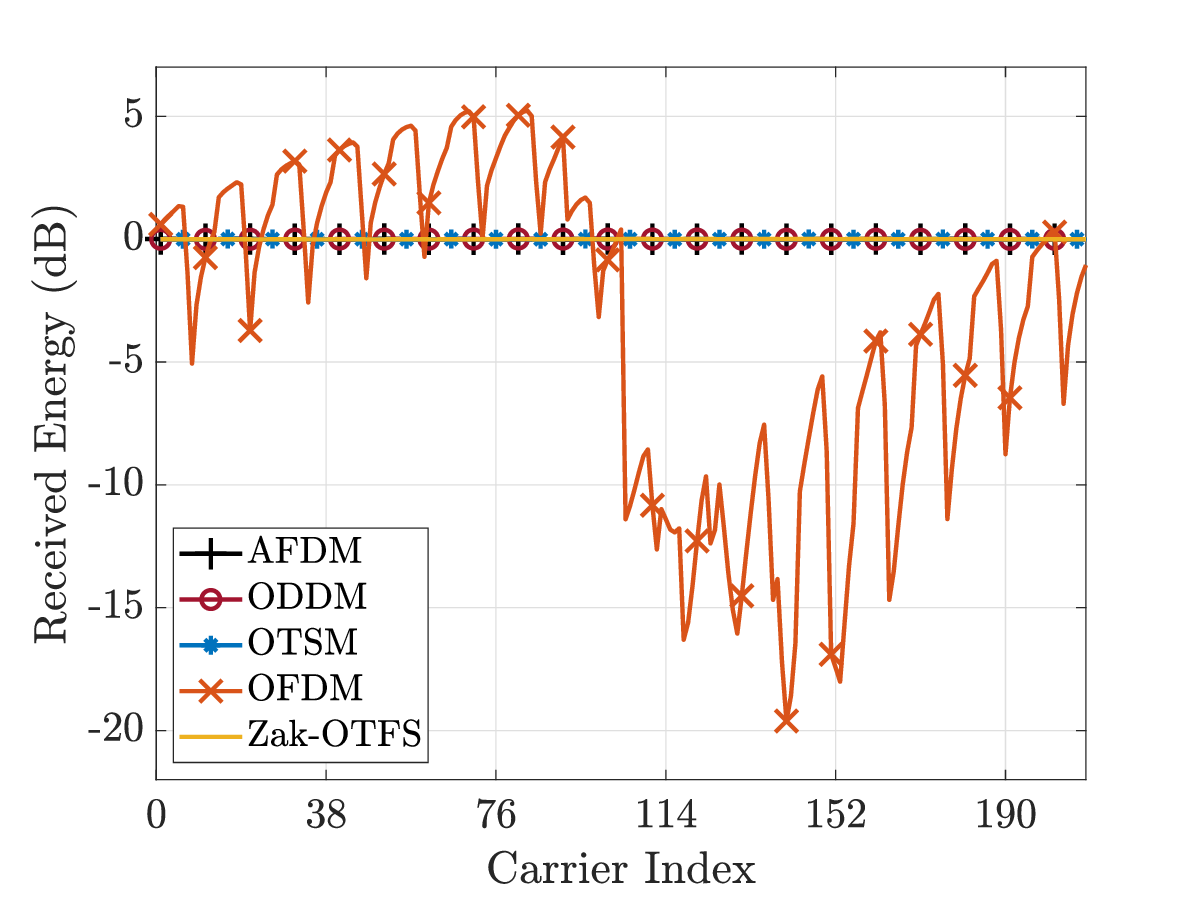}
    \caption{Received energy per-carrier for various modulations.}
        \label{fig:received_energy}
    \end{subfigure}
    \begin{subfigure}{0.43\linewidth}
    \includegraphics[width=\textwidth]{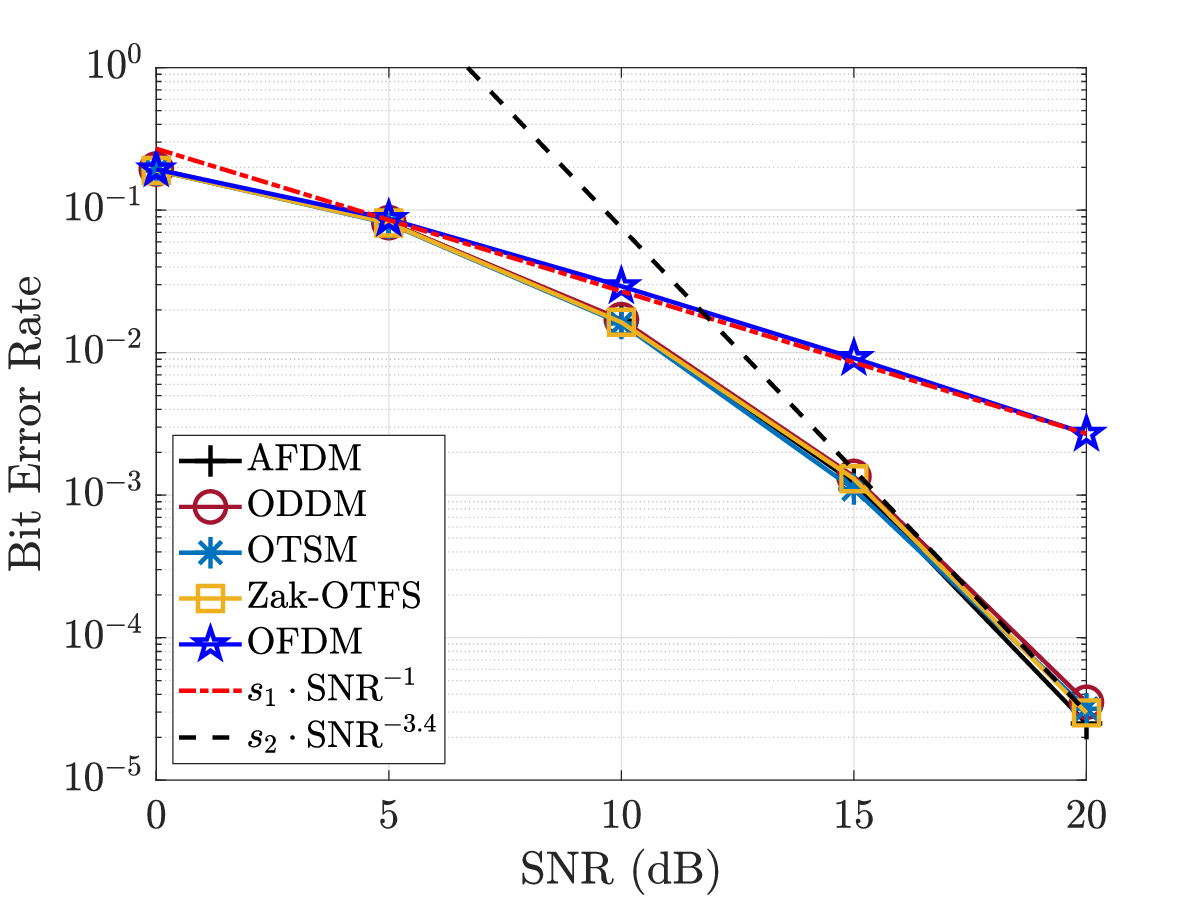}
    \caption{Bit error rate (BER) with perfect channel knowledge.}
        \label{fig:ber_perfectcsi}
    \end{subfigure}
    \caption{\textcolor{black}{(a) Modulation schemes in $\Psi$ -- AFDM, ODDM, OTSM \& Zak-OTFS -- are non-selective with no variation in per-carrier energy, whereas modulations not in $\Psi$ (OFDM) are selective. (b) Modulations in $\Psi$ achieve full delay-Doppler diversity (our channel model has $2-4$ resolvable paths), whereas modulations not in $\Psi$ (OFDM) only achieve a diversity gain of $1$.}} 
    \label{fig:perfectcsi}
\end{figure*}

Following Theorem~\ref{thm:family_pred}, for all modulation schemes in $\Psi$, $\widehat{\mathbf{h}}[k,l] = \mathbf{h}[k,l]$ (and hence $\widehat{\mathbf{G}} = \mathbf{G}$). Hence, the channel estimation and data detection performance of all modulation schemes in $\Psi$ (including the four specific modulations in Section~\ref{subsec:family_complex_hadamard}) is expected to coincide. It has been established in~\cite{Chockalingam2025_divzak} that Zak-OTFS (a member of $\Psi$) achieves full delay-Doppler diversity. Due to performance equivalence, this implies that every modulation in $\Psi$ also achieves full diversity.

\subsection{Application to Multi-Waveform Co-Existence}
\label{subsec:family_multiwvf}

Note that in the examples presented in Section~\ref{subsec:family_complex_hadamard}, the choice of the complex Hadamard matrix $\mathbf{H}_{(i)_{{}_{M}}}$ is fixed for all values of $(i)_{{}_{M}} \in \mathbb{Z}_{M}$. Choosing different complex Hadamard matrices for different values of $(i)_{{}_{M}} \in \mathbb{Z}_{M}$ enables \emph{multiplexing} multiple waveforms in the same transmission. This might find utility in multi-user, multi-service networks where different waveforms may be allocated to different users / services depending on their requirement (e.g., AFDM has lower PAPR than Zak-OTFS but larger inter-waveform interference~\cite{Mehrotra2025_WCLSpread,Mehrotra2025_EURASIP}; hence, AFDM may be preferred in single-user settings vs Zak-OTFS in multi-user settings). 

\begin{lemma}
    \label{lmm:family_orth}
    Choosing $\mathbf{H}_{(i)_{{}_{M}}}$ dependent on $(i)_{{}_{M}} \in \mathbb{Z}_{M}$ does not affect the orthonormality of the resulting modulation basis.
\end{lemma}


\begin{IEEEproof}
    Evaluating the inner product between $\boldsymbol{\phi}_{i}, \boldsymbol{\phi}_{j} \in \Psi$:
    \begin{align*}
        \big\langle \boldsymbol{\phi}_{i}, \boldsymbol{\phi}_{j} \big\rangle\!=\!\frac{1}{N} &\sum_{n \in \mathbb{Z}_{MN}} \mathbf{H}_{(i)_{{}_{M}}}^{*}[\lfloor \nicefrac{n}{M} \rfloor, \lfloor \nicefrac{i}{M} \rfloor] \mathds{1}\big\{n \equiv i \bmod{M}\big\} \nonumber \\ &~~~~\times \mathbf{H}_{(j)_{{}_{M}}}[\lfloor \nicefrac{n}{M} \rfloor, \lfloor \nicefrac{j}{M} \rfloor] \mathds{1}\big\{n \equiv j \bmod{M}\big\}.
    \end{align*}

    The indicator function implies $i \equiv j \bmod{M}$ and $n = (i)_{{}_{M}} + \Delta M,~\Delta \in \mathbb{Z}_{N}$; hence:
    \begin{align*}
        \big\langle \boldsymbol{\phi}_{i}, \boldsymbol{\phi}_{j} \big\rangle &= \frac{1}{N} \sum_{\Delta \in \mathbb{Z}_{N}} \mathbf{H}_{(i)_{{}_{M}}}^{*}[\Delta, \lfloor \nicefrac{i}{M} \rfloor] \mathbf{H}_{(i)_{{}_{M}}}[\Delta, \lfloor \nicefrac{j}{M} \rfloor] \nonumber \\ &~~~~~~~~~~~~\times \mathds{1}\big\{i \equiv j \bmod{M}\big\} \nonumber \\
        &= \mathds{1}\big\{\lfloor \nicefrac{i}{M} \rfloor = \lfloor \nicefrac{j}{M} \rfloor\big\} \mathds{1}\big\{i \equiv j \bmod{M}\big\},
    \end{align*}
    where the final expression follows from Definition~\ref{def:complex_Hadamard}.
\end{IEEEproof}

Theorem~\ref{thm:family_pred} ensures that such multiplexed modulation bases are predictable and non-selective. Therefore, the performance of such multiplexed schemes remains equivalent to schemes where $\mathbf{H}_{(i)_{{}_{M}}}$ is fixed for all $(i)_{{}_{M}} \in \mathbb{Z}_{M}$. Fig.~\ref{fig:txrx} illustrates the transceiver architecture of our system (with waveform multiplexing), assuming $\mathbf{U} = \mathbf{I}_{MN}$ in~\eqref{eq:sys1}. At the transmitter \& receiver,~\eqref{eq:prelim3}, \eqref{eq:prelim4} are implemented for the bases in~\eqref{eq:sys1} as\footnote{\textcolor{black}{\label{ftnt:txrx} In matrix-vector form as $\mathbf{X}[(n)_{{}_{M}},:] = \frac{1}{\sqrt{N}} \mathbf{S}[(n)_{{}_{M}},:] \mathbf{H}_{(n)_{{}_{M}}}^{\top}$ and $\mathbf{R}[(i)_{{}_{M}},:] = \frac{1}{\sqrt{N}} \mathbf{Y}[(i)_{{}_{M}},:] \mathbf{H}_{(i)_{{}_{M}}}^{*}$, where $\mathbf{A} \in \big\{\mathbf{X},\mathbf{S},\mathbf{Y},\mathbf{R}\big\}$ denotes the $M \times N$ matrix representation of an $MN$-length sequence $\mathbf{a} \in \big\{\mathbf{x},\mathbf{s},\mathbf{y},\mathbf{r}\big\}$ with $\mathbf{A}[(n)_{{}_{M}},\lfloor \nicefrac{n}{M} \rfloor] = \mathbf{a}[n]$ and $\mathbf{A}[i,:]$ denoting the $i$th row of $\mathbf{A}$.}}:
\begin{align}
    \label{eq:tx1}
    \mathbf{x}[n] &= \frac{1}{\sqrt{N}} \sum_{\Delta \in \mathbb{Z}_{N}} \mathbf{s}[(n)_{{}_{M}}+\Delta M] \mathbf{H}_{(n)_{{}_{M}}}[\lfloor \nicefrac{n}{M} \rfloor, \Delta], \\
    \label{eq:rx1}
    \mathbf{r}[i] &= \frac{1}{\sqrt{N}} \sum_{\Delta \in \mathbb{Z}_{N}} \mathbf{H}_{(i)_{{}_{M}}}^{*}[\Delta, \lfloor \nicefrac{i}{M} \rfloor] \mathbf{y}[(i)_{{}_{M}}+\Delta M].
\end{align}

The overall hardware complexity is limited by the worst-case complexity of computing the above $N$-length projections. Since projection onto standard $N \times N$ complex Hadamard matrices (DFT, Walsh, etc.) is possible in $O(N\log N)$ complexity, the overall hardware complexity is thus $O(MN\log N)$.

\section{Numerical Results}
\label{sec:results}


To evaluate the theoretical results from Section~\ref{sec:family}, we simulate examples of modulation schemes in $\Psi$ -- Zak-OTFS, ODDM, OTSM \& AFDM -- (see Section~\ref{subsec:family_complex_hadamard}) as well as OFDM (which is not in $\Psi$). We conduct numerical simulations using a 3GPP-compliant $P=6$ path Vehicular-A (Veh-A) channel model~\cite{veh_a}, whose power-delay profile is shown in Table~\ref{tab:veh_a}. The Doppler of each path is simulated as $\nu_i = \nu_{\max}\cos(\theta_i)$, with $\theta_i$ uniformly distributed in $[-\pi, \pi)$ and $\nu_{\max} = 815$ Hz denoting the maximum channel Doppler spread\footnote{Our channel model is representative of real propagation environments since it considers \textit{fractional} delay and Doppler shifts -- the path delays in Table~\ref{tab:veh_a} being non-integer multiples of the delay resolution $\nicefrac{1}{B}$, and the Doppler shifts $\nu_i = \nu_{\max}\cos(\theta_i)$ being non-integer multiples of the Doppler resolution $\nicefrac{1}{T}$.}. We consider parameters: $M=13, N=16$, $B = 0.39$ MHz, $T = 0.533$ ms, and uncoded $4$-QAM (quadrature amplitude modulation) transmissions. We assume Gaussian-sinc pulse shaping as per~\cite{Chockalingam2025_gs}. \textcolor{black}{The channel parameters are chosen to satisfy the condition in Theorem~\ref{thm:family_pred}.} \textcolor{black}{We perform data detection using the minimum mean squared error (MMSE) estimator\footnote{\textcolor{black}{$\widehat{\mathbf{s}} = (\widehat{\mathbf{G}}^{\mathsf{H}}\widehat{\mathbf{G}}+\sigma^2\mathbf{I}_{MN})^{-1} \widehat{\mathbf{G}}^{\mathsf{H}} \mathbf{r}$}}~\cite{Tse2005}.} We do not simulate larger frame sizes and wider bandwidths since we have shown in~\cite{Mehrotra2025_WCLSpread} that the uncoded data detection performance remains similar.

\begin{table}[!t]
    \centering
    \caption{Power-delay profile of Veh-A channel model}
    \begin{tabular}{|c|c|c|c|c|c|c|}
         \hline
         Path index $i$ & $1$ & $2$ & $3$ & $4$ & $5$ & $6$ \\
         \hline
         Delay $\tau_i (\mu s)$ & $0$ & $0.31$ & $0.71$ & $1.09$ & $1.73$ & $2.51$ \\
         \hline
         Relative power (dB) & $0$ & \textcolor{black}{$-1$} & \textcolor{black}{$-9$} & \textcolor{black}{$-10$} & \textcolor{black}{$-15$} & \textcolor{black}{$-20$} \\
         \hline
    \end{tabular}
    \label{tab:veh_a}
\end{table}

\begin{figure*}
    \centering
    \begin{subfigure}{0.43\linewidth}
    \includegraphics[width=\textwidth]{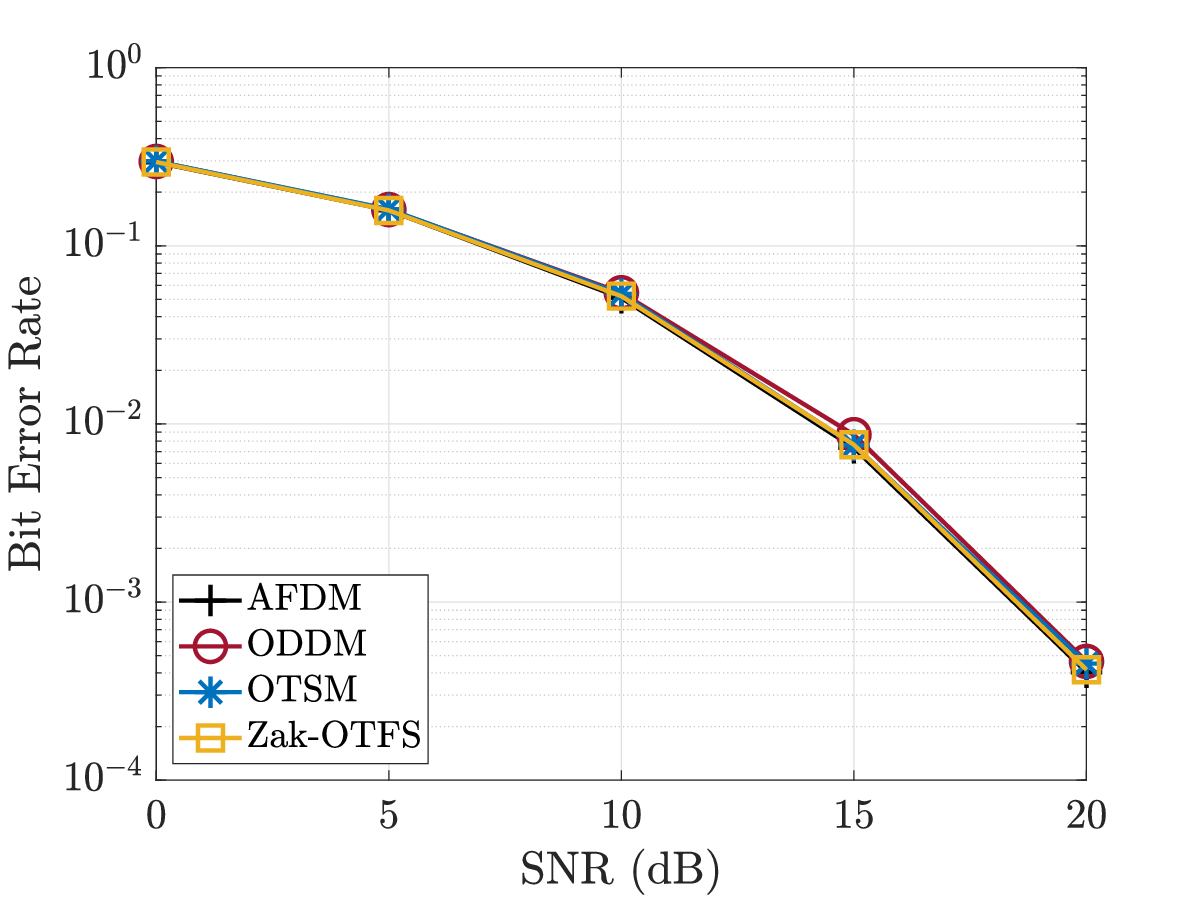}
    \caption{Bit error rate (BER).}
        \label{fig:ber_estcsi}
    \end{subfigure}
    \begin{subfigure}{0.43\linewidth}
    \includegraphics[width=\textwidth]{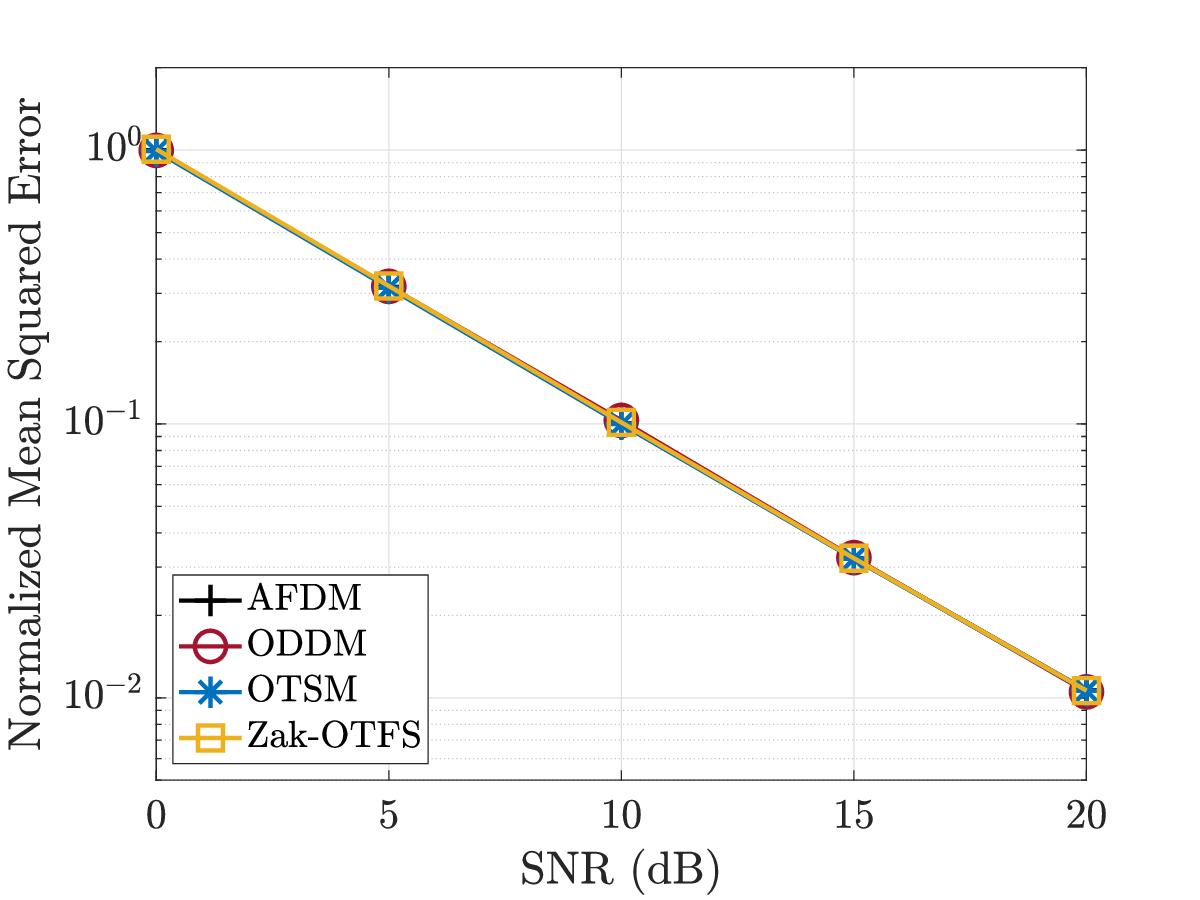}
    \caption{Normalized mean squared error (NMSE).}
        \label{fig:nmse_pdr}
    \end{subfigure}
    \caption{Uncoded $4$-QAM data detection with estimated channel. \textcolor{black}{Modulations in $\Psi$ have equivalent performance.}}
    \label{fig:estcsi}
\end{figure*}


Fig.~\ref{fig:perfectcsi}(\subref{fig:received_energy}) plots the received energy per-carrier for various modulations. Consistent with Theorem~\ref{thm:family_pred}, we observe that the considered modulations in $\Psi$ -- AFDM, ODDM, OTSM \& Zak-OTFS -- exhibit non-selectivity (same received energy per-carrier), whereas OFDM has large per-carrier energy variation.


Fig.~\ref{fig:perfectcsi}(\subref{fig:ber_perfectcsi}) plots the bit error rate (BER) with perfect channel knowledge, i.e., known matrix $\mathbf{G}$ in~\eqref{eq:prelim5}, at the receiver. Fig.~\ref{fig:estcsi} plots the uncoded $4$-QAM BER and normalized mean squared error (NMSE) for channel estimation via~\eqref{eq:prelim6} using a separate pilot frame with pilot signal-to-noise ratio (SNR) equal to the data SNR. \textcolor{black}{Consistent with the findings in Section~\ref{subsec:family_prop}, we observe \emph{equivalent}, full diversity achieving data detection performance for modulations in $\Psi$, whereas modulations not in $\Psi$ (OFDM) only achieve a diversity gain of $1$.} 

\section{Conclusion}
\label{sec:conclusion}

There has been considerable interest in designing new modulation schemes for doubly-selective channels. In this letter, we established that previously proposed modulations for this problem belong to a common waveform family. Under benign channel conditions, all modulations in this waveform family do not undergo symbol fading, allow channel estimation with low overhead, and offer equivalent, full diversity achieving performance. Applications to multi-waveform co-existence via waveform multiplexing were also discussed. Future work will further explore the information theoretic aspects of the proposed framework, and explore system-level applications to problems such as link adaptation, multi-user scheduling, etc.

\bibliographystyle{IEEEtran}
\bibliography{references}
\end{document}